\pdfoutput=1

\documentclass[11pt]{article}

\usepackage[final]{ACL2023}

\usepackage{times}
\usepackage{latexsym}
\usepackage{paralist}
\usepackage{enumitem}
\usepackage{comment}

\usepackage[T1]{fontenc}

\usepackage[utf8]{inputenc}
\usepackage{xspace}

\usepackage{microtype}
\usepackage{amsmath}

\usepackage{inconsolata}
\usepackage{pgfplots}
\usepackage{booktabs}
\usepackage{multirow}
\usepackage{hyperref}

\usetikzlibrary{patterns}
\usetikzlibrary{matrix}
\pgfplotsset{compat=1.7}
\definecolor{colorone}{RGB}{90,180,172}
\definecolor{colortwo}{RGB}{199,234,229}
\definecolor{colorthree}{RGB}{1,102,94}

\definecolor{colorwayone}{RGB}{166,97,26}
\definecolor{colorwaytwo}{RGB}{223,194,125}
\definecolor{colorwaythree}{RGB}{245,245,245}
\definecolor{colorwayfour}{RGB}{128,205,193}
\definecolor{colorwayfive}{RGB}{1,133,113}

\newcommand{\ourname}{DataFinder\xspace}

\definecolor{myblue}{rgb}{0.9, 0.1, 0.94}
\definecolor{mygreen}{rgb}{0.64, 0.56, 0.88}
\definecolor{myyellow}{rgb}{0.98, 0.94, 0.75}
\definecolor{mygreen}{rgb}{0, 1, 0}

\newcommand{\std}[1]{\color{gray}{\xspace#1\xspace}}

\title{\ourname: Scientific Dataset Recommendation from\\Natural Language Descriptions}

\author{
Vijay Viswanathan$^{1}$ \quad
Luyu Gao$^{1}$ \\
\textbf{Tongshuang Wu}$^{1}$\quad
\textbf{Pengfei Liu}$^{2, 3}$\quad 
\textbf{Graham Neubig}$^{1, 3}$ \\
$^1$Carnegie Mellon University \quad
$^2$Shanghai Jiao Tong University \quad 
$^3$Inspired Cognition \\
 {\small \texttt{\{vijayv, luyug, sherryw, gneubig\}@cs.cmu.edu} \quad \texttt{stefanpengfei@gmail.com}}
}

\begin{document}
\maketitle
\begin{abstract}
Modern machine learning relies on datasets to develop and validate research ideas. Given the growth of publicly available data, finding the right dataset to use is increasingly difficult. Any research question imposes explicit and implicit constraints on how well a given dataset will enable researchers to answer this question, such as dataset size, modality, and domain. We operationalize the task of recommending datasets given a short natural language description of a research idea, to help people find relevant datasets for their needs. Dataset recommendation poses unique challenges as an information retrieval problem; datasets are hard to directly index for search and there are no corpora readily available for this task. To facilitate this task, we build  \emph{the DataFinder Dataset} which consists of a larger automatically-constructed training set (17.5K queries) and a smaller expert-annotated evaluation set (392 queries). Using this data, we compare various information retrieval algorithms on our test set and present a superior bi-encoder retriever for text-based dataset recommendation. This system, trained on \emph{the DataFinder Dataset}, finds more relevant search results than existing third-party dataset search engines. To encourage progress on dataset recommendation, we release our dataset and models to the public.%
\footnote{Code and data: \url{https://github.com/viswavi/datafinder}}
\end{abstract}

\section{Introduction}

Innovation in modern machine learning (ML) depends on datasets.
The revolution of neural network models in computer vision \citep{alexnet} was enabled by the ImageNet Large Scale Visual Recognition Challenge \citep{imagenet}. Similarly, data-driven models for syntactic parsing saw rapid development after adopting the Penn Treebank \citep{ptb, palmer2010linguistic}.

\begin{figure}[t]
  \centering
    \includegraphics[trim={1.5cm 12.5cm 13cm 0cm}, clip, width=1.48\linewidth]{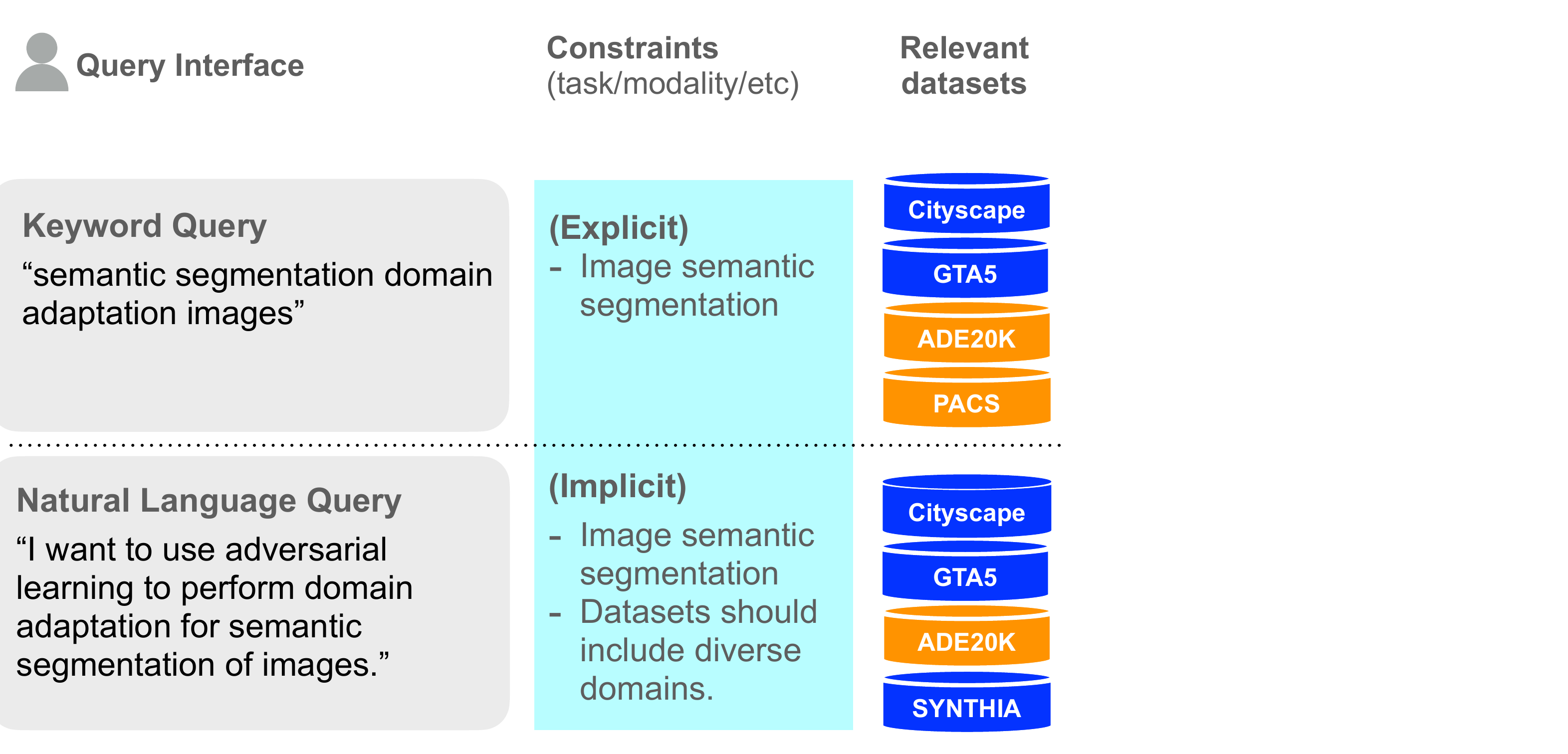}
  \caption{Queries for dataset recommendation impose constraints on the type of dataset desired. Keyword queries make these constraints explicit, while full-sentence queries impose implicit constraints. Ground truth relevant datasets for this query are colored in blue.
  }
\label{fig:case_study_example}
\vspace{-7pt}
\end{figure}

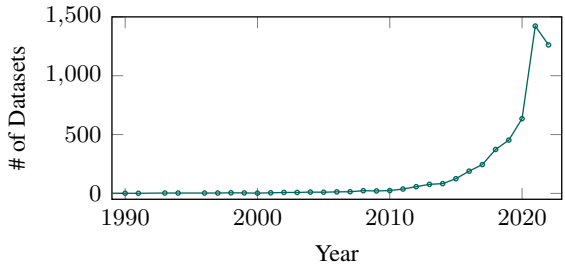
\begin{figure}[t]
\centering
\begin{tikzpicture}
  \begin{axis}[ 
  width=\linewidth,
  line width=0.5,
  grid=major, %
  tick label style={font=\normalsize},
  legend style={nodes={scale=0.4, transform shape}},
  label style={font=\normalsize},
  legend image post style={mark=triangle},
  grid style={white},
  xlabel={Year},
ylabel={\# of Datasets},
   y tick label style={
    font=\small
 },
 ymin=-50,
 ymax=1500,
 xmin=1989,
 xmax=2023,
    x tick label style={font=\small, /pgf/number format/.cd,%
          scaled x ticks = false,
          set thousands separator={}},
    label style={font=\small},
legend style={at={(1,1)}, anchor=north east,  draw=none, fill=none},
    height=4.0cm,
    width=7.5cm,
  ]
    \addplot[colorthree] coordinates
      {(1982, 1) (1987, 1) (1990, 1) (1991, 1) (1993, 3) (1994, 3) (1996, 3) (1997, 3) (1998, 5) (1999, 4) (2000, 2) (2001, 5) (2002, 8) (2003, 8) (2004, 11) (2005, 10) (2006, 13) (2007, 14) (2008, 23) (2009, 21) (2010, 24) (2011, 37) (2012, 57) (2013, 77) (2014, 83) (2015, 125) (2016, 189) (2017, 245) (2018, 373) (2019, 453) (2020, 635) (2021, 1422) (2022, 1262)};;

      \addplot+ [only marks, colorthree, mark=o, mark size=0.8pt] coordinates
      {(1982, 1) (1987, 1) (1990, 1) (1991, 1) (1993, 3) (1994, 3) (1996, 3) (1997, 3) (1998, 5) (1999, 4) (2000, 2) (2001, 5) (2002, 8) (2003, 8) (2004, 11) (2005, 10) (2006, 13) (2007, 14) (2008, 23) (2009, 21) (2010, 24) (2011, 37) (2012, 57) (2013, 77) (2014, 83) (2015, 125) (2016, 189) (2017, 245) (2018, 373) (2019, 453) (2020, 635) (2021, 1422) (2022, 1262)};;

  \end{axis}
\end{tikzpicture}
\vspace{-3em}
\caption{The number of public AI datasets has exploded in recent years. Here we show the \# released from 1990 to 2022 according to Papers with Code.\footnote{\url{https://paperswithcode.com/}}} 
\label{fig:rate_of_datasets}
\vspace{-0.8em}
\end{figure}

With the growth of research in ML and artificial intelligence (AI), there are hundreds of datasets published every year (shown in \autoref{fig:rate_of_datasets}). Knowing which to use for a given research idea can be difficult~\cite{paullada2021data}. To illustrate, consider a real query from a graduate student who says, \emph{``I want to use adversarial learning to perform domain adaptation for semantic segmentation of images.''} They have implicitly issued two requirements: they need a dataset for semantic segmentation of images, and they want datasets that include diverse visual domains. %
A researcher may intuitively select popular, generic semantic segmentation datasets like COCO \citep{lin2014microsoft} or ADE20K \citep{zhou2019semantic}, but these are insufficient to cover the query's requirement of supporting domain adaptation. How can we infer the intent of the researcher and make appropriate recommendations?

To study this problem, we operationalize the task of ``\textbf{dataset recommendation}'': given  \emph{a full-sentence description} or \emph{keywords} describing a research topic, recommend datasets to support research on this topic (\S\ref{sec:task}). A concrete example is shown in \autoref{fig:case_study_example}. 
This task was introduced by \citet{farber2021recommending}, who framed it as text classification. In contrast, we naturally treat this task as retrieval \cite{Manning2005IntroductionTI}, where the \emph{search collection} is a set of datasets represented textually with dataset descriptions\footnote{From \url{www.paperswithcode.com}}, structured metadata, and published ``citances'' --- references from published papers that use each dataset~\citep{Nakov2004CitancesCS}. %
This framework allows us to measure performance with rigorous ranking metrics such as mean reciprocal rank~\citep{radev-etal-2002-evaluating}.

To strengthen evaluation, we build a dataset, \emph{the \ourname Dataset},
to measure how well we can recommend datasets for a given description (\S\ref{sec:dataset}).
As a proxy for real-world queries for our dataset recommendation engine, we construct queries from \emph{paper abstracts} to simulate researchers' historical information needs. We then identify the datasets used in a given paper, either through manual annotations (for our small test set) or using heuristic matching (for our large training set). To our knowledge, this is the first expert-annotated corpus for dataset recommendation, and we believe this can serve as a challenging testbed for researchers interested in representing and searching complex data.

We evaluate three existing ranking algorithms on our dataset and task formation, as a step towards solving this task: BM25 \citep{Robertson2009ThePR}, nearest neighbor retrieval, and dense retrieval with neural bi-encoders~\citep{dpr}. BM-25 is a standard baseline for text search, nearest neighbor retrieval lets us measure the degree to which this task requires generalization to new queries, and bi-encoders are among the most effective search models used today \citep{Zhong2022EvaluatingTA}. Compared with third-party keyword-centric dataset search engines, a bi-encoder model trained on \emph{\ourname}
is far more effective at finding relevant datasets. We show that finetuning the bi-encoder on our training set is crucial for good performance. However, we observe that this model is as effective when trained and tested on keyphrase queries as on full-sentence queries, suggesting that there is room for improvement in automatically understanding full-sentence queries.

\section{Dataset Recommendation Task}
\label{sec:task}
We establish a new task for automatically recommending relevant datasets given a description of a data-driven system. Given a query  $q$ and a set of datasets $D$, retrieve the most relevant subset $R \subset D$ one could use to test the idea described in $q$.
\autoref{fig:case_study_example}
illustrates this with a real query written by a graduate student.

The query $q$ can take two forms: either a keyword query (the predominant interface for dataset search today \citep{Chapman2019DatasetSA}) or a full-sentence description.
Textual descriptions offer a more flexible input to the recommendation system, with the ability to implicitly specify constraints based on what a researcher wants to study, without needing to carefully construct keywords a priori.

\paragraph{Evaluation Metrics}
Our task framing naturally leads to evaluation by information retrieval metrics that estimate search relevance. 
In our experiments, we use four common metrics included in the \texttt{trec\_eval} package,\footnote{\url{https://github.com/usnistgov/trec_eval}. We use the \texttt{-c} flag for the \texttt{trec\_eval} command.} a standard evaluation tool used in the IR community:

\begin{itemize}[itemsep=-.2em,leftmargin=1em,topsep=0.5em]
\item \textbf{P}recision@$k$: The proportion of relevant items in top $k$ retrieved datasets. If P@$k$ is 1, then every retrieved document is valuable.
\item \textbf{R}ecall@$k$: The fraction of relevant items that are retrieved. If R@$k$ is 1, then the search results are comprehensive.
\item Mean Average Precision (\textbf{MAP}): Assuming we have $m$ relevant datasets in total, and $k_i$ is the rank of the $i^{\text{th}}$ relevant dataset, MAP is calculated as $\sum_{i}^{m} \text{P@}k_i / m$ \citep{Manning2005IntroductionTI}. High MAP indicates strong average search quality over all relevant datasets.
\item Mean Reciprocal Rank (\textbf{MRR}):  The average of the inverse of the ranks at which the first relevant item was retrieved. Assuming $R_i$ is the rank of the $i$-th relevant item in the retrieved result, $MRR$ is calculated as $\sum_{i}^{m} R_i / m$. High MRR means a user sees \emph{at least some} relevant datasets early in the search results.
\end{itemize}

\section{\emph{The \ourname Dataset}}
\label{sec:dataset}

\begin{figure}[t]
\begin{center}

    \includegraphics[width=0.48\textwidth]{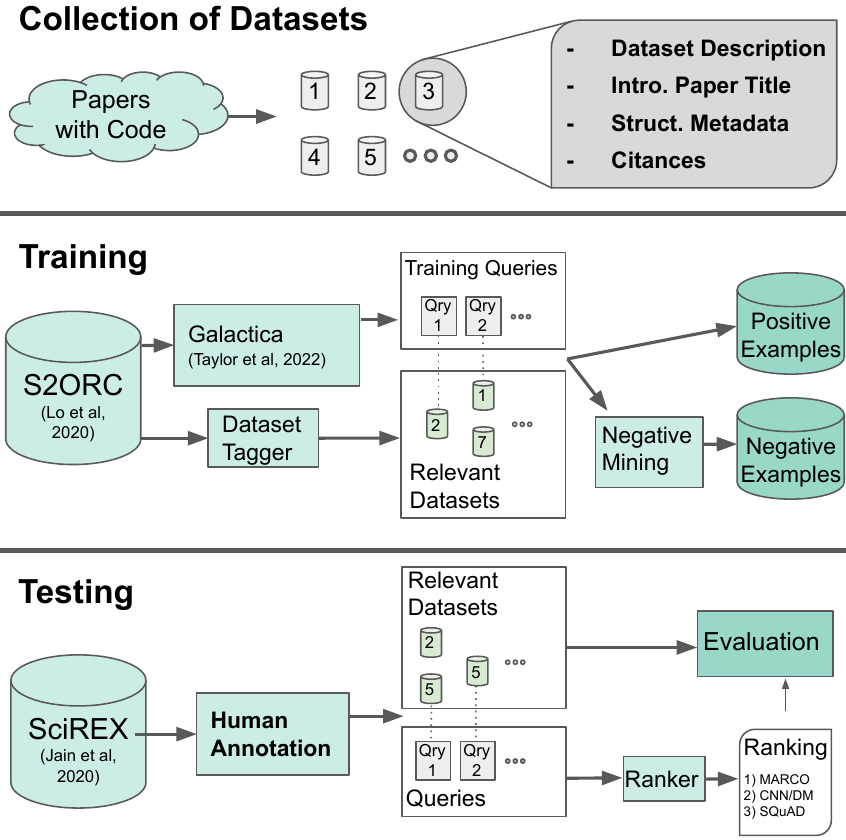}
    
\end{center}
\caption{We search against all datasets on Papers With Code. Our system is trained on a set of simulated queries and target datasets and evaluated on a set of expert-written queries with hand-annotated target datasets. %
}
\label{fig:data_workflow}
\end{figure}

To support this task, we construct a dataset called \emph{The \ourname Dataset} consisting of $(q, R)$ pairs extracted from published English-language scientific proceedings, where each $q$ is either a full-sentence description or a keyword query. 
We collect a large training set through an automated method (for scalability), and we collect a smaller test set using real users' annotations (for reliable and realistic model evaluation).
In both cases, our data collection contains two primary steps: (1) \textbf{collecting search queries} $q$ that a user would use to describe their dataset needs, and (2) \textbf{identifying relevant datasets} $R$ that match the query. Our final training and test sets contain 17495 and 392 queries, respectively.
\autoref{fig:data_workflow} summarizes our data collection approach. We explain the details below and provide further discussion of the limitations of our dataset in the \hyperref[sec:dataset-limitations]{Limitations} section.
We will release our data under a permissive CC-BY License. %

\subsection{Collection of Datasets}
In our task definition, we search over the collection of datasets listed on Papers With Code, a large public index of papers which includes metadata for over 7000 datasets and benchmarks. For most datasets, Papers With Code Datasets stores a short human-written dataset description, a list of different names used to refer to the dataset (known as ``variants''), and structured metadata such as the year released, the number of papers reported as using the dataset, the tasks contained, and the the modality of data.
Many datasets also include the paper that introduced the dataset. We used the dataset description, structured metadata, and the introducing paper's title to textually represent each dataset, and we analyze this design decision in \S\ref{sec:analysis}.

\subsection{Training Set Construction}

To ensure scalability for the training set, we rely on a large corpus of scientific papers, S2ORC \citep{lo-etal-2020-s2orc}. We extract nearly 20,000 abstracts from AI papers that use datasets. To overcome the high cost of manually-annotating queries or relevant datasets, we instead simulate annotations with few-shot-learning and rule-based methods.

\paragraph{Query Collection}
We extract queries from paper abstracts because, intuitively, an abstract will contain the most salient characteristics behind a research idea or contribution. As a result, it is an ideal source for comprehensively collecting potential implicit constraints as shown in \autoref{fig:case_study_example}. 

We simulate query collection with the 6.7B parameter version of Galactica \citep{taylor2022galactica}, a large scientific language model that supports few-shot learning.  In our prompt, we give the model an abstract and ask it to first extract five keyphrases: the tasks mentioned in paper, the task domain of the paper (e.g., biomedical or aerial), the modality of data required, the language of data or labels required, and the length of text required (sentence-level, paragraph-level, or none mentioned). We then ask Galactica to generate a full query containing any salient keyphrases.  We perform few-shot learning using 3 examples in the prompt to guide the model. Our prompt is shown in \autoref{sec:prompt-appendix}.

\paragraph{Relevant Datasets}
\label{sec:training_set_labels}
For our training set, relevant datasets are automatically labeled using the body text of a paper.\footnote{Note that our queries are obtained from the abstract alone while the relevance judgements are obtained from the text body, to encourage more general queries.} We apply a rule-based procedure to identify the dataset used in a given paper (corresponding to an abstract whose query has been auto-labeled). For each paper, we tag all datasets that satisfy two conditions: the paper must cite the paper that introduces the dataset, and the paper must mention the dataset by name twice.\footnote{We apply the additional requirement that the counted dataset mentions must occur in a section with section title containing ``results'', ``experiment'', ``evaluation'', ``result'', ``training'', or ``testing'', to avoid non-salient dataset mentions, such as those commonly occurring in ``related work".
}

This tagging procedure is restrictive and emphasizes precision (i.e., an identified dataset is indeed used in the paper) over recall (i.e., all the used datasets are identified). Nonetheless, using this procedure, we tag 17,495 papers from S2ORC with at least one dataset from our collection of datasets.

To estimate the quality of these tagged labels, we manually examined 200 tagged paper-dataset pairs. Each pair was labeled as correct if the paper authors would have realistically had to download the dataset in order to write the paper.
92.5\% (185/200) of dataset tags were deemed correct. 

\subsection{Test Set Construction}
\label{sec:test_set}

To accurately approximate how humans might search for datasets, we employed AI researchers and practitioners to annotate our test set. As mentioned above, the dataset collection requires both \emph{query collection} and \emph{relevant dataset collection}.
We use SciREX \citep{jain-etal-2020-scirex}, a human-annotated set of 438 full-text papers from major AI venues originally developed for research into full-text information extraction,  as the basis of our test set. We choose this dataset because it naturally supports our dataset collection described below.

\paragraph{Query Collection}
We collect search queries by asking annotators to digest, extract, and rephrase key information in research paper abstracts.

\emph{Annotators.} To ensure domain expertise, we recruited 27 students, faculty, and recent alumni of graduate programs in machine learning, computer vision, robotics, NLP, and statistics from major US universities. We recruited 23 annotators on a voluntary basis through word of mouth; for the rest, we offered 10 USD in compensation. We sent each annotator a Google Form that contained between 10 and 20 abstracts to annotate. The instructions provided for that form are shown in \autoref{sec:annotations-appendix}.

\emph{Annotation structure.} For each abstract, we asked annotators to extract metadata regarding the abstract's task, domain, modality, language of data required, and length of data required. These metadata serve as \textbf{keyphrase queries}.
Then, based on these keyphrases, we also ask the annotator to write a sentence that best reflects the dataset need of the given paper/abstract, which becomes the \emph{full-sentence query}.
Qualitatively, we found that the keyphrases helped annotators better ground and concretize their queries, and the queries often contain (a subset of) these keyphrases.

\emph{Model assistance.} To encourage more efficient labeling \citep{Wang2021WantTR}, we provided auto-suggestions for each field from GPT-3 \citep{gpt3} and Galactica 6.7B \citep{taylor2022galactica} to help annotators. We note that annotators rarely applied these suggestions directly --- annotators accepted the final full-sentence query generated by either large language model only 7\% of the time.

\paragraph{Relevant Datasets}
For each paper, SciREX contains annotations for mentions of all ``salient" datasets, defined as datasets that ``take part in the results of the article'' \citep{jain-etal-2020-scirex}. We used these annotations as initial suggestions for the datasets used in each paper. The authors of this paper then skimmed all 438 papers in SciREX and noted the datasets used in each paper. 46 papers were omitted because they either used datasets not listed on Papers With Code or were purely theory-based papers with no relevant datasets, leaving a final set of 392 test examples.

We double-annotated 10 papers with the datasets used. The annotators labeled the exact same set of datasets for 8 out of 10 papers, with a Fleiss-Davies kappa of 0.667, suggesting that inter-annotator agreement for our ``relevant dataset'' annotations is substantial \citep{davies_fleiss, nltk}.

\subsection{Dataset Analysis}
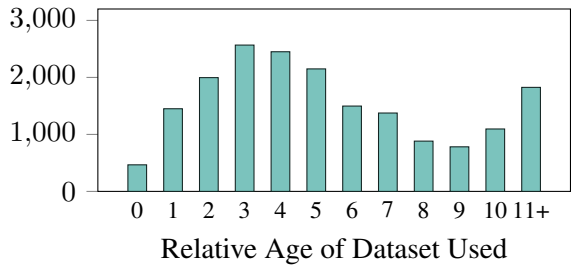
\begin{figure}[t]
\centering
    \begin{tikzpicture}
        \begin{axis}[
            ybar,
            ymin=0,
            bar width=7pt, 
            xlabel={Relative Age of Dataset Used},
            symbolic x coords={0,1,2,3,4,5,6,7,8,9,10,11+},
            xtick=data,
            x tick label style={font=\small},
            legend cell align={left},
            xlabel near ticks,
            ylabel near ticks,
            ytick align=outside,
            ytick pos=left,
            xtick align=inside,
            xtick style={draw=none},
            ylabel shift=-3 pt,
            height=4.0cm,
            width=7.8cm,
            ymax=3200,
            xtick distance=1,
            ]
            \addplot[colorone!20!black,fill=colorone!80!white] coordinates {

    (0, 466)
    (1, 1450)
    (2, 1996)
    (3, 2567)
    (4, 2450)
    (5, 2149)
    (6, 1497)
    (7, 1375)
    (8, 882)
    (9, 782)
    (10, 1095)
    (11+, 1825)
};
        \end{axis}
        
    \end{tikzpicture}
    \caption{Distribution of relative year of datasets used across all papers that used a dataset.}
    \label{fig:dataset_recency}
\vspace{-10pt}
\end{figure}

Using this set of paper-dataset tags, what can we learn about how researchers use datasets? %

    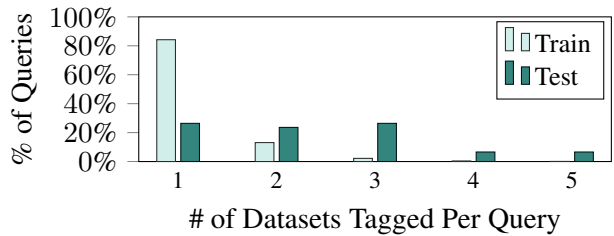
\begin{figure}[t]
\centering
       \begin{tikzpicture}
        \begin{axis}[
            ybar,
            ymin=0,
            bar width=7pt, 
            xlabel={\# of Datasets Tagged Per Query},
            ylabel={\% of Queries},
            symbolic x coords={0,1,2,3,4,5},
            xtick=data,
            x tick label style={font=\small},
            legend cell align={left},
            xlabel near ticks,
            ylabel near ticks,
            ytick align=outside,
            ytick pos=left,
            xtick align=inside,
            xtick style={draw=none},
            ytick align=outside,
            ytick pos=left,
            yticklabel={\pgfmathprintnumber\tick\%},
            ylabel shift=-3 pt,
            height=3.5cm,
            width=7.8cm,
            ymax=100,
            legend style={at={(0.985,0.97)},anchor=north east,nodes={scale=0.9, transform shape}},
            xtick distance=1,
            ]
            \addplot[colortwo!20!black,fill=colortwo!80!white] coordinates {
                (1, 84.2)
                (2, 13.1)
                (3, 2.2)
                (4, 0.4)
                (5, 0.1)
            };
            
            \addplot[colorthree!20!black,fill=colorthree!80!white] coordinates {
                (1, 26.4)
                (2, 23.6)
                (3, 26.4)
                (4, 6.6)
                (5, 6.6)
            };

            \legend{Train, Test}

        \end{axis}
        
    \end{tikzpicture}
    \vspace{-30pt}
    \caption{The distribution of the number of datasets tagged in each paper, in train and test sets}
    \label{fig:dataset_tags_per_instance}
\end{figure}

Our final collected dataset contains 17,495 training queries and 392 test queries. 
The training examples usually associate queries with a single dataset much more frequently than our test set does. This is due to our rule-based tagging scheme, which emphasizes precise labels over recall. Meanwhile, the median query from our expert-annotated test set had 3 relevant datasets associated with it. 
We also observed interesting dataset usage patterns: 
\begin{itemize}[itemsep=-.2em,leftmargin=1em,topsep=0.5em]
\item \textbf{Researchers tend to converge towards popular datasets.} Analyzing dataset usage by community,\footnote{We define ``communities'' by publication venues: \emph{ACL}, \emph{EMNLP}, \emph{NAACL}, \emph{TACL}, \emph{COLING} for NLP, \emph{CVPR}, \emph{ICCV}, \emph{WACV} for Vision, \emph{IROS}, \emph{ICRA}, \emph{IJRR} for Robotics, and \emph{NeurIPS}, \emph{ICML} \emph{ICLR} for Machine Learning. We include proceedings from associated workshops in each community.} we find that in all fields, among all papers that use some publicly available dataset, more than 50\% papers in our training set use at least one of the top-5 most popular datasets. Most surprisingly, nearly half of the papers tagged in the robotics community use the KITTI dataset \citep{Geiger2013VisionMR}. 
\item \textbf{Researchers tend to rely on recent datasets.}

    \begin{figure*}[t]
\centering
        \begin{tikzpicture}
        \begin{axis}[
            ybar,
            ymin=0,
            enlarge x limits=0.08,
            ylabel={\% of Papers},
            symbolic x coords={SQuAD, SNLI, COCO, SST, GLUE, MultiNLI, UD, SICK},
            xtick=data,
            nodes near coords,
            every node near coord/.append style={font=\small},
            nodes near coords align={vertical},
            x tick label style={rotate=45, anchor=north east, inner sep=0mm, font=\small},
            legend cell align={left},
            xlabel near ticks,
            ylabel near ticks,
            ytick align=outside,
            ytick pos=left,
            yticklabel={\pgfmathprintnumber\tick\%},
            xtick align=inside,
            xtick style={draw=none},
            ylabel shift=-3 pt,
            height=3.5cm,
            width=7.8cm,
            ymin=0,
            ymax=60,
            xtick distance=1,
            ]
            \addplot[colorone!20!black,fill=colorone!80!white] coordinates {
    (SQuAD, 16.3)
    (SNLI, 12.7)
    (COCO, 8.8)
    (SST, 7.5)
    (GLUE, 5.9)
    (MultiNLI, 4.4)
    (UD, 4.1)
    (SICK, 3.3)
};

\node[] at (axis cs: UD,45) {NLP};
        \end{axis}
        
    \end{tikzpicture}     \hspace{4pt} \raisebox{-0.37em}{\begin{tikzpicture}
        \begin{axis}[
            ybar,
            ymin=0,
            enlarge x limits=0.08,
            symbolic x coords={COCO, ImageNet, KITTI, CIFAR-10, UCF101, CelebA, Cityscapes, MPII},
            xtick=data,
            nodes near coords,
            every node near coord/.append style={font=\small},
            nodes near coords align={vertical},
            x tick label style={rotate=45, anchor=north east, inner sep=0mm, font=\small},
            legend cell align={left},
            xlabel near ticks,
            ylabel near ticks,
            ymajorticks=false,
            xtick align=inside,
            xtick style={draw=none},
            ylabel shift=-3 pt,
            height=3.5cm,
            width=7.8cm,
            ymin=0,
            ymax=60,
            xtick distance=1,
            ]
            \addplot[colorone!20!black,fill=colorone!80!white] coordinates {
    (COCO, 20.2)
    (ImageNet, 13.9)
    (KITTI, 8.4)
    (CIFAR-10, 6.5)
    (UCF101, 5.2)
    (CelebA, 5.0)
    (Cityscapes, 4.1)
    (MPII, 2.8)
};
\node[] at (axis cs: Cityscapes,45) {CV};

        \end{axis}
        
    \end{tikzpicture}}

    \hspace{1.5pt}\begin{tikzpicture}
        \begin{axis}[
            ybar,
            ymin=0,
            enlarge x limits=0.08,
            ylabel={\% of Papers},
            symbolic x coords={KITTI, MuJoCo, COCO, ImageNet, Cityscapes, ScanNet, ShapeNet, CARLA},
            xtick=data,
            nodes near coords,
            every node near coord/.append style={font=\small},
            nodes near coords align={vertical},
            x tick label style={rotate=45, anchor=north east, inner sep=0mm, font=\small},
            legend cell align={left},
            xlabel near ticks,
            ylabel near ticks,
            ytick align=outside,
            ytick pos=left,
            yticklabel={\pgfmathprintnumber\tick\%},
            xtick align=inside,
            xtick style={draw=none},
            ylabel shift=-3 pt,
            height=3.5cm,
            width=7.8cm,
            ymin=0,
            ymax=60,
            xtick distance=1,
            ]
            \addplot[colorone!20!black,fill=colorone!80!white] coordinates {

    (KITTI, 44.9)
    (MuJoCo, 11.8)
    (COCO, 8.8)
    (ImageNet, 7.4)
    (Cityscapes, 4.4)
    (ScanNet, 4.0)
    (ShapeNet, 3.3)
    (CARLA, 2.9)
};
\node[] at (axis cs: ShapeNet,45) {Robotics};
        \end{axis}
        
    \end{tikzpicture}     \hspace{-1pt}\raisebox{-0.02em}{\begin{tikzpicture}
        \begin{axis}[
            ybar,
            ymin=0,
            enlarge x limits=0.08,
            symbolic x coords={CIFAR-10, ImageNet, CelebA, COCO, MuJoCo, SQuAD, MovieLens, SNLI},
            xtick=data,
            nodes near coords,
            every node near coord/.append style={font=\small},
            nodes near coords align={vertical},
            x tick label style={rotate=45, anchor=north east, inner sep=0mm, font=\small},
            legend cell align={left},
            xlabel near ticks,
            ylabel near ticks,
            ymajorticks=false,
            xtick align=inside,
            xtick style={draw=none},
            ylabel shift=-3 pt,
            height=3.5cm,
            width=7.8cm,
            ymin=0,
            ymax=60,
            xtick distance=1,
            ]
            \addplot[colorone!20!black,fill=colorone!80!white] coordinates {

    (CIFAR-10, 28.9)
    (ImageNet, 18.4)
    (CelebA, 11.5)
    (COCO, 7.9)
    (MuJoCo, 7.2)
    (SQuAD, 2.4)
    (MovieLens, 2.3)
    (SNLI, 2.0)
};
\node[] at (axis cs: MovieLens,45) {ML};

        \end{axis}
        
    \end{tikzpicture}}
    \vspace{-30pt}
    \caption{We analyze the distribution of datasets used in NLP, robotics, vision, and machine learning research.}
    \label{fig:datasets_by_genre}
\end{figure*}
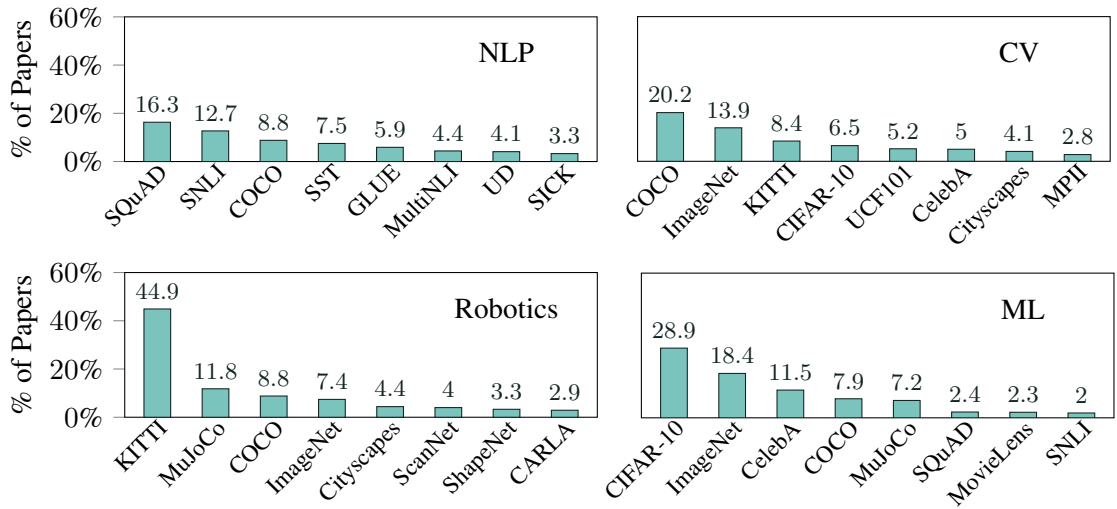

In \autoref{fig:dataset_recency}, we see the distribution of relative ages of datasets used (i.e., the year between when a dataset is published, and when a corresponding paper uses it for experiments). In \autoref{fig:dataset_recency}, We observe that the average dataset used by a paper was released 5 years before the paper's publication (with a median of 5.6 years), but we also see a significant long tail of older datasets. This means that while some papers use traditional datasets, most papers exclusively use recently published datasets. %
\end{itemize}

These patterns hint that researchers might overlook less cited datasets that match their needs in favor of standard \emph{status-quo} datasets. This motivates the need for nuanced dataset recommendation.

\section{Experimental Setup on \emph{DataFinder}}
\label{sec:methods}
How do popular methods perform on our new task and new dataset? How does our new paradigm differ from existing commercial search engines? In this section, we describe a set of standard methods which we benchmark, and we consider which third-party search engines to use for comparison.

\subsection{Task Framing}
\label{subsec:task_framing_and_eval}
We formulate dataset recommendation as a ranking task. Given a query $q$ and a search corpus of datasets $D$, rank the datasets $d \in D$ based on a query-dataset similarity function $\text{sim}(q, d)$ and return the top $k$ datasets. We compare three ways of defining $\text{sim}(q, d)$: term-based retrieval, nearest-neighbor retrieval, and neural retrieval.

\subsection{Models to Benchmark}

To retrieve datasets for a query, we find the nearest datasets to that query in a vector space. We represent each query and dataset in a vector space using three different approaches:

\paragraph{Term-Based Retrieval}
We evaluated a BM25 retriever for this task, since this is a standard baseline algorithm for information retrieval. We implement BM25 \citep{Robertson1999OkapiKeenbowAT} using Pyserini \citep{pyserini}.\footnote{We run BM25 with $k_1=0.8$ and $b=0.4$.}

\paragraph{Nearest-Neighbor Retrieval}
To understand the extent to which this task requires generalization to new queries unseen at training time, we experiment with direct $k$-nearest-neighbor retrieval against the training set. For a new query, we identify the most similar queries in the training set and return the relevant datasets from these training set examples. In other words, each dataset is represented by vectors corresponding to all training set queries attached to that dataset. In practice we investigate two types of feature extractors: TF-IDF \citep{Jones2004ASI} and SciBERT~\citep{Beltagy2019SciBERT}.

\paragraph{Neural Retrieval}
We implement a bi-encoder retriever using the Tevatron package.\footnote{\url{https://github.com/texttron/tevatron}} In this framework, we encode each query and document into a shared vector space and estimate similarity via the inner product between query and document vectors. We represent each document with the \verb+BERT+ embedding \citep{devlin-etal-2019-bert} of its [CLS] token:
$$
    \text{sim}(q, d) = \text{cls}(\verb+BERT+(q))^T \text{cls}(\verb+BERT+(d))
$$
where $\text{cls}(\cdot)$ denotes the operation of accessing the [CLS] token representation from the contextual encoding \citep{Gao2021RethinkTO}.
For retrieval, we separately encode all queries and documents and retrieve using efficient similarity search.
Following recent work \citep{dpr}, we minimize a contrastive loss and select hard negatives using BM25 for training. We initialize the bi-encoder with SciBERT~\citep{Beltagy2019SciBERT} and finetune it on our training set. This model takes 20 minutes to finetune on one 11GB Nvidia GPU.

\subsection{Comparison with Search Engines}
\label{sec:commercial_search_engines_method}
Besides benchmarking existing methods, we also compare the methods enabled by our new data recommendation task against the standard paradigm for dataset search --- to use a conventional search engine with short queries \citep{Kacprzak2019CharacterisingDS}. 
We measured the performance of third-party dataset search engines taking as input either keyword queries or full-sentence method descriptions.

We compare on our test set with two third-party systems-- \emph{Google Dataset Search}\footnote{\url{https://datasetsearch.research.google.com}} \citep{brickley2019google} and \emph{Papers with Code}\footnote{\url{https://paperswithcode.com/datasets}} search. Google Dataset Search supports a large dataset collection, so we limit results to those from Papers with Code to allow comparison with the ground truth.

Our test set annotators frequently entered multiple keyphrases for each keyphrase type (e.g. ``question answering, recognizing textual entailment'' for the Task field). We constructed multiple queries by taking the Cartesian product of each set of keyphrases from each field, deduplicating tokens that occurred multiple times in each query. After running each query against a commercial search engine, results were combined using balanced interleaving \citep{Joachims2002OptimizingSE}.

\section{Evaluation}

\subsection{Time Filtering}
The queries in our test set were made from papers published between 2012 and 2020\footnote{We could not include more recent papers in our query construction process, because SciREX was released in 2020.}, with median year 2017. In contrast, half the datasets in our search corpus were introduced in 2018 or later.
To account for this discrepancy, for each query $q$, we only rank the subset of datasets $D' = \{d \in D \text{ } | \text{ year}(d) \leq \text{year}(q)\}$ that were introduced in the same year or earlier than the query.

\subsection{Benchmarking and Comparisons}

\begin{table}[t] 
  \centering
  \renewcommand{\arraystretch}{1.1}
  \fontsize{9}{9.5}\selectfont
  \setlength{\tabcolsep}{2.5pt}
  \begin{tabular}{@{} r|rrrr @{} }
  \toprule
   \textbf{Model} & \textbf{P@5} & \textbf{R@5} & \textbf{MAP} & \textbf{MRR} \\  \midrule\midrule
    \multicolumn{4}{l}{Full-Sentence Queries} \\  \midrule
  BM25  & 4.7\std{$\pm0.1$} & 11.6\std{$\pm1.7$} & 8.0\std{$\pm\ 1.3$} & 14.5\std{$\pm2.0$} \\
  kNN (TF-IDF) & 5.5\std{$\pm$0.6} & 12.3\std{$\pm$1.6} & 7.8\std{$\pm$1.1} & 15.5\std{$\pm$2.0}  \\
  kNN (BERT)  & 7.1\std{$\pm$0.7} & 14.2\std{$\pm$1.5} & 9.7\std{$\pm$1.2} & 21.3\std{$\pm$2.3}  \\
  Bi-Encoder & \textbf{16.0}\std{$\pm$1.1} & \textbf{31.2}\std{$\pm$2.2}  & \textbf{23.4}\std{$\pm$1.9}  & \textbf{42.6}\std{$\pm$2.7}  \\
\midrule
    \multicolumn{4}{l}{Keyphrase Queries} \\  \midrule
  BM25  & 6.6\std{$\pm$0.5} & 15.3\std{$\pm$1.1}  & 11.4\std{$\pm$0.8} & 19.9\std{$\pm$1.5}  \\
  kNN (TF-IDF) & 2.7\std{$\pm$0.4} & 5.9\std{$\pm$1.1} & 3.3\std{$\pm$0.7} & 8.2\std{$\pm$1.6}  \\
  kNN (BERT)  & 2.8\std{$\pm$0.4} & 5.8\std{$\pm$1.1} &   3.3\std{$\pm$1.1} & 7.3\std{$\pm$1.3}  \\
  Bi-Encoder & \textbf{16.5}\std{$\pm$1.0} & \textbf{32.4}\std{$\pm$2.2} & \textbf{23.3}\std{$\pm$1.8} & \textbf{42.3}\std{$\pm$2.6}   \\ 
  \bottomrule
  \end{tabular}
  \caption{A comparison of methods on full-sentence and keyword search shows that the neural bi-encoder performs best by a significant margin. Standard deviations are obtained via bootstrap sampling on the test set.
  }
  \label{table:standard_metric_results} 
\end{table}

\begin{table}[t] 
  \renewcommand{\arraystretch}{0.9}
  \centering
  \small
  \begin{tabular}{r|rrrr}
  \toprule
   \textbf{Model} & \textbf{P@5} & \textbf{R@5} & \textbf{MAP} & \textbf{MRR} \\  \midrule\midrule

  PwC (\emph{descriptions}) & 0.6 & 1.7  & 0.9 & 1.2  \\
  PwC (\emph{keywords})   & 3.5  & 10.0 & 6.5 & 9.1  \\
  \midrule
  Google (\emph{descriptions})  & 0.1 & 0.1 & 0.1 & 0.3  \\
  Google  (\emph{keywords})  & 9.7 & 19.5 & 12.3 & 24.0  \\
  \midrule
  Ours (\emph{descriptions})  & 16.0 & 31.2 & 23.4 & 42.6   \\ 
  Ours (\emph{keywords})  & 16.5 & 32.4 & 23.3 & 42.3    \\ 
  
  \bottomrule
  \end{tabular}
  \caption{Comparing third-party search engines (\emph{Papers with Code} and \emph{Google Dataset Search}) against our \ourname system using a bi-encoder architecture.
  }
  \label{table:comparing_against_commercial_engines} 
\end{table}

\paragraph{Benchmarking shows that \ourname benefits from deep semantic matching.}
In \autoref{table:standard_metric_results}, we report retrieval metrics on the methods described in \S\ref{sec:methods}. To determine the standard deviation of each metric, we use bootstrap resampling \citep{Koehn2004StatisticalST} over all test set queries. Term-based retrieval (BM25) performs poorly in this setting, while the neural bi-encoder model excels. This suggests our task requires capturing semantic similarity beyond what term matching can provide. Term-based kNN search is not effective, implying that generalization to new queries is necessary for this task.

\paragraph{Commercial Search Engines are not effective on \emph{\ourname}.}
\label{sec:comparing_with_engines}

In \autoref{table:comparing_against_commercial_engines}, we compare our proposed retrieval system against third-party dataset search engines. For each search engine, we choose the top 5 results before computing metrics. 

We find these third-party search engines do not effectively support full-sentence queries. We speculate these search engines are adapted from term-based web search engines.
In contrast, our neural retriever gives much better search results using both keyword search and full-sentence query search.

\begin{figure}[t]
    \centering
    \includegraphics[width=0.48\textwidth]{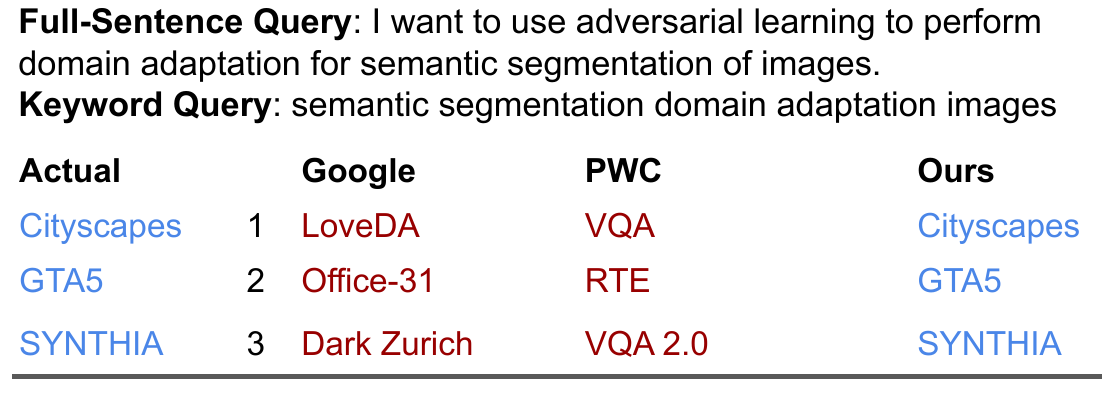}
    
    \includegraphics[width=0.48\textwidth]{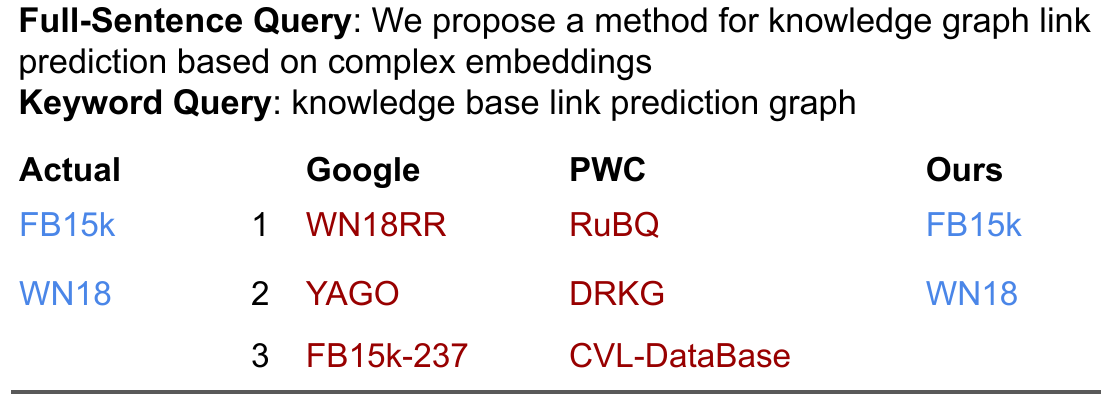}
    
    \includegraphics[width=0.48\textwidth]{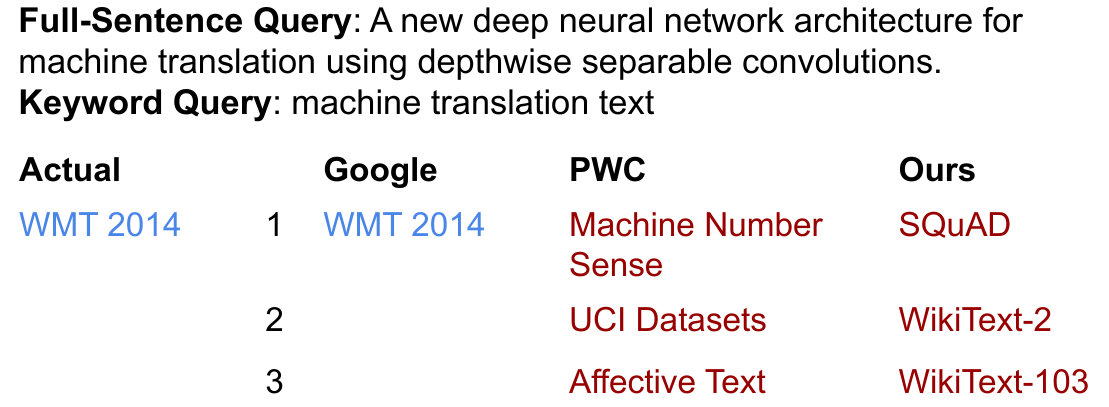}
    \vspace{-1em}
    \caption{We qualitatively compare the retrieval behavior of a neural biencoder retriever (trained on \emph{\ourname}) and third-party dataset search engines.
    }
    \label{fig:qualitative_results}
\vspace{-5pt}
\end{figure}

\subsection{Qualitative Analysis}
Examples in \autoref{fig:qualitative_results} highlight the tradeoffs between third-party search engines and models trained on \ourname.
In the first two examples, we see keyword-based search engines struggle when dealing with terms that could apply to many datasets, such as ``semantic segmentation'' or ``link prediction''. These keywords offer a limited specification on the relevant dataset, but a system trained on simulated search queries from real papers can learn implicit filters expressed in a query.

On the final example, our system incorrectly focuses on the deep architecture described (``deep neural network architecture [...] using depthwise separable convolutions'') rather than the task described by the user (``machine translation''). Improving query understanding for long queries is a key opportunity for improvement on this dataset.

\subsection{More In-depth Exploration}
\label{sec:analysis}

We perform in-depth qualitative analyses to understand the trade-offs of different query formats and dataset representations.

\paragraph{Comparing full-sentence vs keyword queries}
As mentioned above, we compare two versions of the \ourname-based system: one trained and tested with description queries and the other with keyword queries. We observe that using keyword queries offers similar performance to using full-sentence descriptions for dataset search. This suggests more work should be done on making better use of implicit requirements in full-sentence descriptions for natural language dataset search.

\paragraph{Key factors for successful queries}
What information in queries is most important for effective dataset retrieval? Using human-annotated keyphrase queries in our test set, we experiment with concealing particular information from the keyphrase query. %

    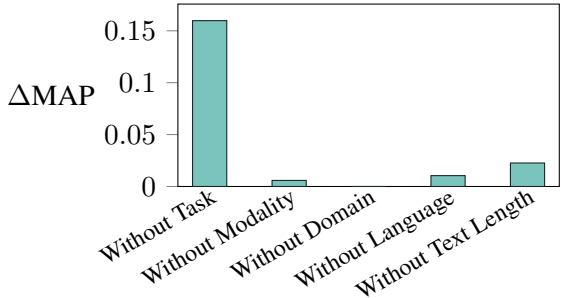
\begin{figure}[t]
\centering
    \begin{tikzpicture}
        \begin{axis}[
            ybar,
            ymin=0,
            bar width=12.8pt, 
            symbolic x coords={Without Task, Without Modality, Without Domain, Without Language, Without Text Length},
            xtick=data,
            x tick label style={rotate=30, anchor=north east, inner sep=0mm, font=\small},
            legend cell align={left},
            ylabel={$\Delta$MAP},
            xlabel near ticks,
            ylabel near ticks,
            ylabel style={rotate=-90},
            ytick align=outside,
            ytick pos=left,
            xtick align=inside,
            xtick style={draw=none},
            ylabel shift=-4 pt,
            height=4.0cm,
            width=6.6cm,
            yticklabel style={
                    /pgf/number format/fixed,
                    /pgf/number format/precision=2
            },
            xtick distance=1,
            ]
            \addplot[colorone!20!black,fill=colorone!80!white] coordinates {

    (Without Task, 0.159781)
    (Without Modality, 0.005841)
    (Without Domain, 0.000)
    (Without Language, 0.010458)
    (Without Text Length, 0.022628)
};
        \end{axis}
        
    \end{tikzpicture}
    \caption{Comparison of the reduction in the MAP metric of the retrieval results after removing different types of query terms (e.g. keywords related to the task or language the researcher is interested in studying).}
    \label{fig:query_ablation}
\end{figure}

In \autoref{fig:query_ablation}, we see task information is critical for dataset search; removing task keywords from queries reduces MAP from 23.5 to 7.5 (statistically significant with $p <0.001$ by a paired bootstrap t-test).  Removing constraints on the language of text data also causes a significant drop in MAP ($p<0.0001$). Removing keywords for text length causes an insignificant reduction in MAP ($p = 0.15$), though it causes a statistically significant reduction on other metrics not shown in \autoref{fig:query_ablation}: P@5 and R@5.
Based on inspection of our test set, we speculate that domain keywords are unnecessary because the domain is typically implied by task keywords.

\paragraph{Comparing textual representations of datasets}
\label{sec:dataset-representations}

\begin{table}[t] 
  \centering
  \renewcommand{\arraystretch}{1.1}
  \fontsize{9}{9.5}\selectfont
  \setlength{\tabcolsep}{2.5pt}
  \begin{tabular}{@{} r|rrrr @{} }
  \toprule
   \textbf{Model} & \textbf{P@5} & \textbf{R@5} & \textbf{MAP} & \textbf{MRR} \\  \midrule\midrule
    \multicolumn{4}{l}{Full-Sentence Queries} \\  \midrule
  Description  & 15.3\std{$\pm1.0$} & 30.0\std{$\pm2.1$} & 23.0\std{$\pm\ 1.9$} & 42.8\std{$\pm2.7$} \\
  + Struct. Info & \textbf{16.0}\std{$\pm$1.1} & \textbf{31.2}\std{$\pm$2.2} & \textbf{23.3}\std{$\pm$1.8} & \textbf{42.4}\std{$\pm$2.7}  \\
  + Citances  & 15.8\std{$\pm$1.1} & 30.8\std{$\pm$2.2} & 23.1\std{$\pm$1.9} & 42.2\std{$\pm$2.7}  \\
\midrule
    \multicolumn{4}{l}{Keyphrase Queries} \\  \midrule
   Description & 13.1\std{$\pm$1.0} & 25.6\std{$\pm$2.0}  & 17.4\std{$\pm$1.6} & 33.1\std{$\pm$2.5}  \\
 + Struct. Info & 16.6\std{$\pm$1.1} & 32.7\std{$\pm$2.2} & 23.5\std{$\pm$1.8} & 42.8\std{$\pm$2.8}  \\
  + Citances  & \textbf{16.8}\std{$\pm$1.0} & \textbf{33.4}\std{$\pm$2.2} &  \textbf{23.6}\std{$\pm$1.8} & \textbf{43.0}\std{$\pm$2.6}  \\
  \bottomrule
  \end{tabular}
  \caption{Adding structured metadata for each dataset's textual representation significantly improves keyphrase search quality using a neural bi-encoder. We compute standard deviations via bootstrap resampling. We use the "Description + Struct. Info" textual representation for all other experiments in this paper.
  }
  \label{table:comparing_dataset_representations} 
\end{table}

We represent datasets textually with a community-generated dataset description from PapersWithCode, along with the title of the paper that introduced the dataset. We experiment with enriching this dataset representation in two ways. We first add structured metadata about each dataset (e.g. tasks, modality, number of papers that use each dataset on PapersWithCode). We cumulatively experiment with adding citances --- sentences from other papers around a citation --- to capture how others use the dataset. In \autoref{table:comparing_dataset_representations}, our neural bi-encoder achieves similar retrieval performance on all 3 representations for full-sentence search.

Keyword search is more sensitive to  dataset representation. adding structured information to the dataset representation provides significant benefits for keyword search. This suggests keyword search requires more specific dataset metadata than full-sentence search does to be effective.

\paragraph{The value of finetuning}
\label{sec:finetuning-ablation}

Our bi-encoder retriever is finetuned on our training set. Given the effort required to construct a training set for tasks like dataset recommendation, is this step necessary?

\begin{table}[t] 
  \centering
  \renewcommand{\arraystretch}{1.1}
  \fontsize{9}{9.5}\selectfont
  \setlength{\tabcolsep}{2.5pt}
  \begin{tabular}{@{} r|rrrr @{} }
  \toprule
   \textbf{Model} & \textbf{P@5} & \textbf{R@5} & \textbf{MAP} & \textbf{MRR} \\  \midrule\midrule
    \multicolumn{4}{l}{Full-Sentence Queries} \\  \midrule
  SciBERT (finetuned)  & \textbf{16.0} & \textbf{31.2} & \textbf{23.3} & \textbf{42.4}  \\
  SciBERT (not finetuned) & 0.0 & 0.0 & 0.0 & 0.0  \\
 COCO-DR (not finetuned)  & 6.1 & 14.8 & 8.8 & 15.7 \\
\midrule
    \multicolumn{4}{l}{Keyphrase Queries} \\  \midrule
   SciBERT (finetuned) & \textbf{16.6} & \textbf{32.7} & \textbf{23.5} & \textbf{42.8}  \\
 SciBERT (not finetuned) & 0.0 & 0.0 & 0.0 & 0.0  \\
 COCO-DR (not finetuned)  & 6.2 & 13.9 &  9.6 & 16.8  \\
  \bottomrule
  \end{tabular}
  \caption{Finetuning for the dataset recommendation task significantly outperforms strong retrieval architectures finetuned for general search, like COCO-DR.}
  \label{table:comparing_finetuning} 
  \vspace{-1pt}
\end{table}

In \autoref{table:comparing_finetuning}, we see that an off-the-shelf SciBERT encoder is ineffective. We observe that our queries, which are abstract descriptions of the user's information need \citep{Ravfogel2023RetrievingTB}, are very far from any documents in the embedding space, making comparison difficult. Using a state-of-the-art encoder, \texttt{COCO-DR Base} --- which is trained for general-purpose passage retrieval on MS MARCO \citep{msmarco}, helps with this issue but still cannot make up for task-specific finetuning.

\section{Related Work}

Most work on scientific dataset recommendation uses traditional search methods, including term-based keyword search and tag search \citep{Lu2012ADS, Kunze2013DatasetR, Sansone2017DATSTD, Chapman2019DatasetSA, brickley2019google, Lhoest2021DatasetsAC}. In 2019, Google Research launched \emph{Dataset Search} \citep{brickley2019google}, offering access to over 2 million public datasets. Our work considers the subset of datasets from their search corpus that have been posted on Papers with Code.

Some work has explored other forms of dataset recommendation.  \citet{Ellefi2016DatasetRF} study using ``source datasets'' as a search query, while \citet{Altaf2019DatasetRV} use a set of related research papers as the user's query.  \citet{farber2021recommending} are the only prior work we are aware of that explores natural language queries for dataset recommendation. They model this task as classification, while we operationalize it as open-domain retrieval. Their dataset uses  abstracts and citation contexts to simulate queries, while we use realistic short queries (with an expert-annotated test set).

\section{Conclusion}

We study the task of dataset recommendation from natural language queries. Our dataset supports search by either full-sentence or keyword queries, but we find that neural search algorithms trained for traditional keyword search are competitive with the same architectures trained for our proposed full-sentence search. An exciting future direction will be to make better use of natural language queries. We release our datasets along with our ranking systems to the public. We hope to spur the community to work on this task or on other tasks that can leverage the summaries, keyphrases, and relevance judgment annotations in our dataset.

\section*{Limitations}
\label{sec:dataset-limitations}
The primary limitations concern the dataset we created, which serves as the foundation of our findings. Our dataset suffers from four key limitations:

\textbf{Reliance on Papers With Code}
Our system is trained and evaluated to retrieve datasets from Papers With Code Datasets (PwC). Unfortunately, PwC is not exhaustive. Several queries in our test set corresponded to datasets that are not in PwC, such as IWSLT 2014 \citep{iwslt_2014}, PASCAL VOC 2010 \citep{pascal_voc_2010}, and CHiME-4 \citep{chime_4_1}. Papers With Code Datasets also skews the publication year of papers used in the \emph{DataFinder Dataset} towards the present (the median years of papers in our train and test set are 2018 and 2017, respectively). For the most part, PwC only includes datasets used by another paper listed in Papers With Code, leading to the systematic omission of datasets seldom used today.

\textbf{Popular dataset bias in the test set}
Our test set is derived from the SciREX corpus \citep{jain-etal-2020-scirex}. This corpus is biased towards popular or influential works: the median number of citations of a paper in SciREX is 129, compared to 19 for any computer science paper in S2ORC. The queries in our test set are therefore more likely to describe mainstream ideas in popular subields of AI.

\textbf{Automatic tagging}
Our training data is generated automatically using a list of canonical dataset names from Papers With Code. This tagger mislabels papers where a dataset is used but never referred to by one of these canonical names (e.g. non-standard abbreviations or capitalizations). Therefore, our training data is noisy and imperfect.

\textbf{Queries in English only}
All queries in our training and test datasets were in English. Therefore, these datasets only support the development of dataset recommendation systems for English-language users. This is a serious limitation, as AI research is increasingly done in languages other English, such as Chinese \citep{cset}.

\section*{Ethics Statement}
Our work has the promise of improving the scientific method in artificial intelligence research, with the particular potential of being useful for younger researchers or students. We built our dataset and search systems with the intention that others could deploy and iterate on our dataset recommendation framework. However, we note that our initial dataset recommendation systems have the potential to increase inequities in two ways.

First, as mentioned in \hyperref[sec:dataset-limitations]{Limitations}, our dataset does not support queries in languages other than English, which may exacerbate inequities in dataset access. We hope future researchers will consider the construction of multilingual dataset search queries as an area for future work.

Second, further study is required to understand how dataset recommendation systems affect the tasks, domains, and datasets that researchers choose to work on. Machine learning models are liable to amplify biases in training data \citep{Hall2022ASS}, and inequities in which domains or tasks receive research attention could have societal consequences. We ask researchers to consider these implications when conducting work on our dataset.

\section*{Acknowledgements}
This work was supported in part by funding from NEC Research Laboratories, DSTA Singapore, the National Science Foundation (NSF) grant IIS-1815528, and a gift from Google.
We thank Sireesh Gururaja, Soham Tiwari, Amanda Bertsch, Liangze Li, Jeremiah Millbauer, Jared Fernandez, Nikhil Angad Bakshi, Bharadwaj Ramachandran, G. Austin Russell, and Siddhant Arora for helping with data collection. We give particular thanks to Carolyn Ros\'e, Saujas Vaduguru, and Ji Min Mun for their helpful discussions and feedback.

\bibliography{anthology,custom}
\bibliographystyle{acl_natbib}

\appendix

\section{Few-Shot Prompt for Generating Keyphrases and Queries}
\label{sec:prompt-appendix}
When constructing our training set, we use in-context few-shot learning with the 6.7B parameter version of Galactica \citep{taylor2022galactica}. We perform in-context few-shot learning with the following prompt:\\
\\
\emph{Given an abstract from an artificial intelligence paper:\\
1) Extract keyphrases regarding the task (e.g. image classification), data modality (e.g. images or speech), domain (e.g. biomedical or aerial), training style (unsupervised, semi-supervised, fully supervised, or reinforcement learning), text length (sentence-level or paragraph-level), language required (e.g. English)\\
2) Write a brief, single-sentence summary containing these relevant keyphrases. This summary must describe the task studied in the paper.\\
\\
Abstract:\\
We study automatic question generation for sentences from text passages in reading comprehension. We introduce an attention-based sequence learning model for the task and investigate the effect of encoding sentence- vs. paragraph-level information. In contrast to all previous work, our model does not rely on hand-crafted rules or a sophisticated NLP pipeline; it is instead trainable end-to-end via sequence-to-sequence learning. Automatic evaluation results show that our system significantly outperforms the state-of-the-art rule-based system. In human evaluations, questions generated by our system are also rated as being more natural (i.e., grammaticality, fluency) and as more difficult to answer (in terms of syntactic and lexical divergence from the original text and reasoning needed to answer).\\
\\
Output: (Task | Modality | Domain | Training Style | Text Length |  Language Required | Single-Sentence Summary)\\
Task: question generation\\
Modality: text\\
Domain: N/A\\
Training Style: fully supervised\\
Text Length: paragraph-level\\
Language Required: N/A\\
Single-Sentence Summary: We propose an improved end-to-end system for automatic question generation.\\
--\\
\\
Abstract:\\
We present a self-supervised approach to estimate flow in camera image and top-view grid map sequences using fully convolutional neural networks in the domain of automated driving. We extend existing approaches for self-supervised optical flow estimation by adding a regularizer expressing motion consistency assuming a static environment. However, as this assumption is violated for other moving traffic participants we also estimate a mask to scale this regularization. Adding a regularization towards motion consistency improves convergence and flow estimation accuracy. Furthermore, we scale the errors due to spatial flow inconsistency by a mask that we derive from the motion mask. This improves accuracy in regions where the flow drastically changes due to a better separation between static and dynamic environment. We apply our approach to optical flow estimation from camera image sequences, validate on odometry estimation and suggest a method to iteratively increase optical flow estimation accuracy using the generated motion masks. Finally, we provide quantitative and qualitative results based on the KITTI odometry and tracking benchmark for scene flow estimation based on grid map sequences. We show that we can improve accuracy and convergence when applying motion and spatial consistency regularization.\\
\\
Output: (Task | Modality | Domain | Training Style | Text Length |  Language Required | Single-Sentence Summary)\\
Task: optical flow estimation\\
Modality: images and top-view grid map sequences\\
Domain: autonomous driving\\
Training Style: unsupervised\\
Text Length: N/A\\
Language Required: N/A\\
Single-Sentence Summary: A system for self-supervised optical flow estimation from images and top-down maps.\\
--\\
\\
Abstract:\\
In this paper, we study the actor-action semantic segmentation problem, which requires joint labeling of both actor and action categories in video frames. One major challenge for this task is that when an actor performs an action, different body parts of the actor provide different types of cues for the action category and may receive inconsistent action labeling when they are labeled independently. To address this issue, we propose an end-to-end region-based actor-action segmentation approach which relies on region masks from an instance segmentation algorithm. Our main novelty is to avoid labeling pixels in a region mask independently - instead we assign a single action label to these pixels to achieve consistent action labeling. When a pixel belongs to multiple region masks, max pooling is applied to resolve labeling conflicts. Our approach uses a two-stream network as the front-end (which learns features capturing both appearance and motion information), and uses two region-based segmentation networks as the back-end (which takes the fused features from the two-stream network as the input and predicts actor-action labeling). Experiments on the A2D dataset demonstrate that both the region-based segmentation strategy and the fused features from the two-stream network contribute to the performance improvements. The proposed approach outperforms the state-of-the-art results by more than 8%
\\
Output: (Task | Modality | Domain | Training Style | Text Length |  Language Required | Single-Sentence Summary)\\
Task: actor-action semantic segmentation\\
Modality: video\\
Domain: N/A\\
Training Style: fully supervised\\
Text Length: N/A\\
Language Required: N/A\\
Single-Sentence Summary: I want to train a supervised model for actor-action semantic segmentation from video.\\
--\\}

For a given abstract that we want to process, we then add this abstract's text to this prompt and ask the language model to generate at most 250 new tokens.

\section{Information on Expert Annotations}
\label{sec:annotations-appendix}

As mentioned in \S\ref{sec:dataset}, we recruited 27 graduate students, faculty, and recent graduate program alumni for our annotation collection process. For each annotator, we received their verbal or written interest in participating in our data collection.

\begin{figure*}[t]
  \centering
    \includegraphics[scale=0.6]{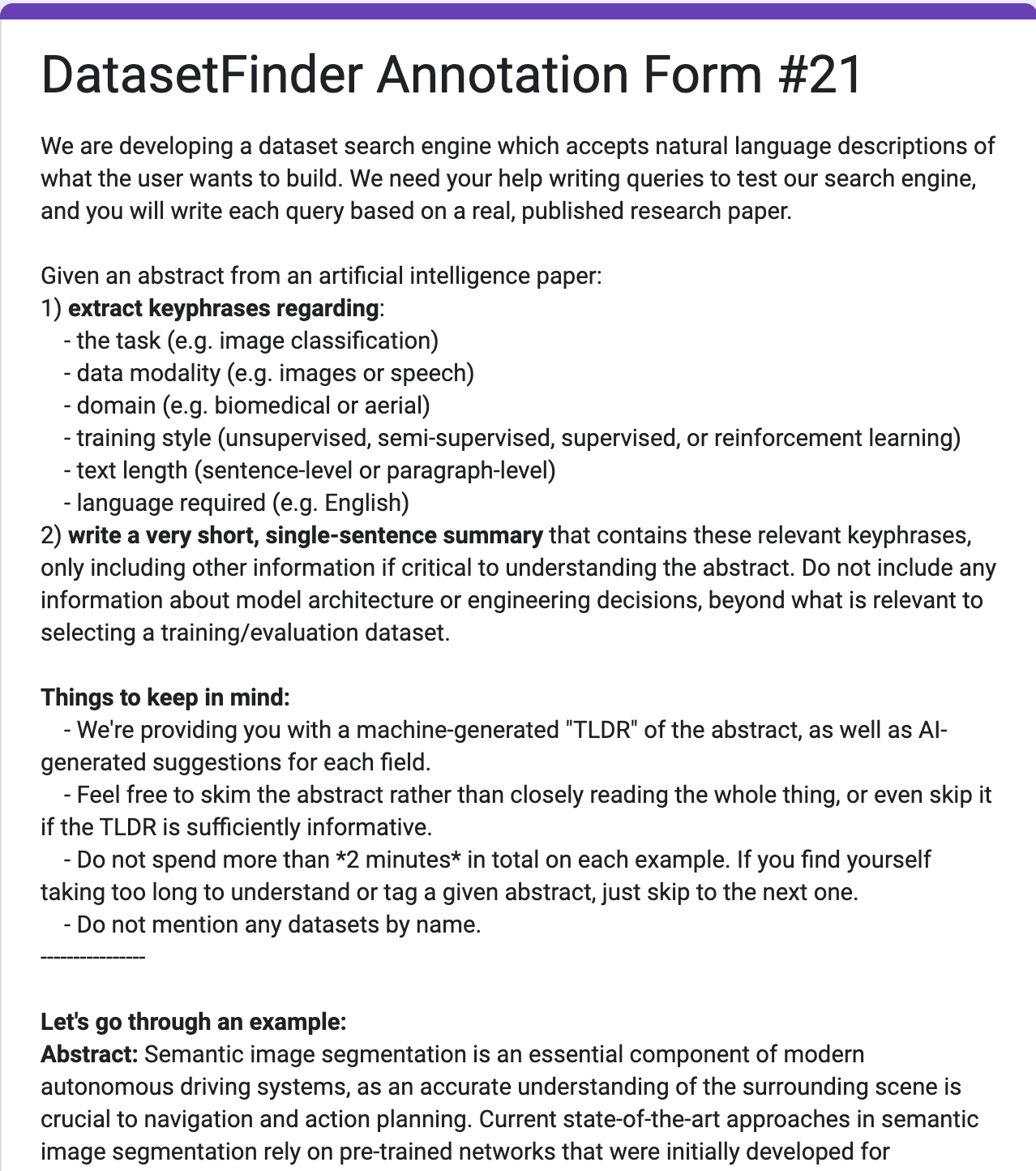}
  \caption{Annotators each annotated 10-20 abstracts for our label collection using a Google Form with the instructions shown here..
  }
\label{fig:annotation-form}
\vspace{-7pt}
\end{figure*}

We then sent them a Google Form containing between 10 and 20 abstracts to annotate. An example of the form instructions is included in \autoref{fig:annotation-form}.

We originally had annotators label the ``Training Style'' (unsupervised, semi-supervised, supervised, or reinforcement learning), in addition to Task, Modality, Domain, Text Length, and Language Required. However, this field saw excessively noisy labels so we ignore this field for our experiments.
\end{document}